\documentclass[%
 reprint,
showpacs,preprintnumbers,
 amsmath,amssymb,
 aps,
]{revtex4-1}
\usepackage{graphicx}
\usepackage{dcolumn}
\usepackage{bm}
\usepackage{epstopdf}

\begin{document}

\preprint{APS/123-QED}

\title{Beyond the thermal model in relativistic heavy-ion collisions}

\author{Georg Wolschin}%
 \email{g.wolschin@thphys.uni-heidelberg.de}
\affiliation{%
 Institut f{\"ur} Theoretische 
Physik
der Universit{\"a}t Heidelberg, Philosophenweg 16, D-69120 Heidelberg, Germany, EU\\
}%


\date{\today}

\begin{abstract}
Deviations from thermal distribution functions of produced particles in relativistic heavy-ion collisions are discussed as indicators for nonequilibrium processes. 
The focus is on rapidity distributions of produced charged hadrons as functions of collision energy and centrality which are used to infer the fraction of  
particles produced from a central fireball as compared to the one from the fragmentation sources that are out of equilibrium with the rest of the system. 
Overall thermal equilibrium would only be reached for large times $t \rightarrow \infty$.



\end{abstract}

\maketitle
\section{Introduction}
\label{intro}

The statistical hadronization or thermal model \cite{hag65} with a limiting temperature $T_{\text{H}}$ has
been successfully used to reproduce, over the full energy range where data have been measured, the ratios of
particle production yields for various hadron species in $\mathrm{e^+ e^-}$, pp and relativistic heavy-ion collisions, e.g. \cite{pbm95,mabe08,pbm16}. 
However, a necessary and sufficient condition for attaining thermal equilibrium in particle 
collisions is provided by the agreement of measured distribution functions with thermal distributions, rather than particle yields.

An example for a thermal distribution may be found in the cosmic microwave background radiation. It has a blackbody spectrum with a temperature of 2.735 K at redshift zero \cite{cobe90}, although there are spatial temperature anisotropies on the level of less than 1 part in $10^4$ which give rise to structure formation, and have meanwhile been measured with excellent accuracy by e.g. the WMAP \cite{wmap09} and Planck \cite{planck15} collaborations.

In relativistic heavy-ion collisions, the distributions of both transverse momentum $p_\mathrm{T}$ as well as rapidity $y$ (or pseudorapidity $\eta$) of produced charged hadrons clearly deviate from thermal distributions. At RHIC and LHC energies, the deviations in a $p_\mathrm{T}$-region of  $0.5$\,GeV/$c\,\lesssim p_\mathrm{T}\,\lesssim7\,\text{GeV}/c$ and the ensuing transition from exponential to power-law $p_\mathrm{T}$-distributions are usually attributed to collective expansion and nonequilibrium processes. 
Above $\sim 7$ GeV/$c$, hard events become visible
which require a pQCD treatment. When integrated over $p_\mathrm{T}$ to obtain particle yields, their contribution is negligible, but decisive as an indicator for nonequilibrium events.

Traces of nonequilibrium behaviour can be found in (pseudo)rapidity distributions of produced charged hadrons as measured by ALICE in PbPb \cite{abb13}, and by ~ATLAS, ALICE
and LHCb in pPb collisions \cite{adam15,atlas13}. The distribution functions have important contributions from the fragmentation regions that are clearly visible in net-proton rapidity distributions at SPS and RHIC energies \cite{bea04,mtw09,dur14}, but also contribute to charged-hadron production. For produced particles, they are found to increase in particle content proportional to $\ln(s_{NN})$ and are not in equilibrium with particles produced in the midrapidity source that arises essentially from low-$x$ gluons.

In the following section several indications for non-thermal system properties found in transverse momentum distributions of produced charged hadrons are reviewed.
Since it turns out that $p_\text{T}$-distributions are, however, inadequate to differentiate fragmentation and central fireball contributions, this serves as a motivation for the investigation of rapidity distributions where the respective role of these individual sources is more obvious. The relevance of the fragmentation contributions is reconsidered in section III, followed by the discussion of pseudorapidity distributions at RHIC and LHC energies with emphasis on the equilibration of the three sources in section IV. The conclusions are drawn in the last section.

\section{Transverse momentum distributions}
Starting from a purely thermal model for particle production in relativistic heavy-ion collisions, the transverse momentum distribution of produced charged hadrons may be 
represented by a relativistic generalization of the Maxwell-Boltzmann distribution that accounts for the fact that the velocity of light $c$ is an upper limit.
The corresponding distribution function was first derived by J\"uttner  \cite{juett11} and is therefore
called the Maxwell-Juettner distribution
\begin{equation}
f(p_\mathrm{T})=\frac{1}{4\pi m^2 T K_2(m/T)}\exp{[-\frac{\gamma(p_\mathrm{T})m}{T}]}
\label{juett}
\end{equation}
with the modified Bessel function of the second kind $K_2(m/T)$, the Lorentz-factor
\begin{equation}
\gamma(p_\mathrm{T})=\sqrt{1+(p_\mathrm{T}/m)^2},
\end{equation}
freeze-out temperature $T$ and 
hadron mass $m$. Here I take $T \equiv T_\mathrm{F} = 120$ MeV (without considering collective expansion, which would lead to a larger effective value), and $m \equiv\langle m \rangle$ as an average value of the masses of pions, kaons and nucleons with contributions of 83\%, 13\% and 4\%, respectively, that correspond to particle production yields in 2.76 TeV PbPb \cite{rgw12}.

\begin{figure}[tph]

\includegraphics[width=9.0cm]{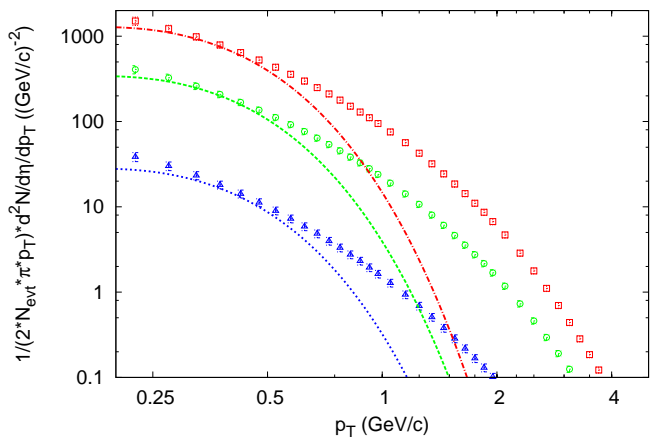}
\caption{(Color online) J\"uttner distribution functions of a relativistic thermal
Maxwell gas at temperature $T$ without collective expansion compared to measured transverse momentum distributions of produced charged hadrons in 2.76~TeV PbPb collisions from ALICE \cite{abe13} for 0--5\%, 30--40\% and 70--80\% centralities (top to bottom; $|\eta| < 0.8$), see text \cite{kw14}.} 
\label{fig1}
\end{figure}

This thermal distribution function is compared in Fig.~\ref{fig1} \cite{kw14} with the charged-hadron distribution measured by ALICE \cite{abe13}  in 2.76 TeV PbPb
for three centralities.
Here the absolute value of the distribution has been adjusted to the measured result at 0--5\%, whereas the normalizations of the calculated distributions at 30--40\% 
and 70--80\%
are obtained from the corresponding ratios of the midrapidity yields \cite{aamo11}. 

Obviously the relativistic thermal distributions fit the measured ones only for very small transverse momenta $p_\mathrm{T} \lesssim 0.5$ GeV/$c$. The generally accepted explanation for this failure is that the system expands collectively. The expansion may to some extent be accounted for phenomenologically by a higher effective temperature $T^* = T + m\, \langle v_\mathrm{T} \rangle^2$ \cite{bea78}. 

The mean transverse velocity $ \langle v_\mathrm{T}\rangle$ depends on the transverse temperature profile and the corresponding velocity at freeze-out time, which are both functions of centrality and may be calculated hydrodynamically, yielding an effective temperature of 
$T^* \simeq 260$ MeV for 0--5\% centrality and a correspondingly broader transverse momentum distribution which agrees with the experimental values in a mean $p_\mathrm{T}$-range.

To treat the transverse expansion in detail, numerous theoretical approaches are available starting from the blast-wave model \cite{sira79} and its boost-invariant generalization \cite{ssh93}. More recently advanced hydrodynamical models such as the ones reviewed in \cite{hesne13,gale13,koda16}
 provide a rather complete description of the collective expansion phase.

The $p_\mathrm{T}$-distributions clearly show a transition from an exponential behaviour in the thermal regime Eq.\,(\ref{juett}) to a power-law behaviour in the $p_\mathrm{T}$-range that is attributed mostly to the recombination of soft partons, and fragmentation of hard partons. 
In addition to detailed theoretical approaches, this transition can be modelled phenomenologically using distribution functions of the form 
\begin{equation}
f(p_\mathrm{T})\propto [1 + (q-1)\, m_\mathrm{T}/T]^{1/(1-q)}
\label{qfct}
\end{equation}
with the transverse mass $m_\mathrm{T}=\sqrt{m^2 + p_\mathrm{T}^2}$, a freezeout temperature $T$ and a dimensionless parameter $q \gtrsim 1$. For $q \rightarrow 1$ the exponential distribution
(extensive statistics) is 
recovered, whereas $q>1$ may be fitted to the measured distribution functions.
 
The functional form (\ref{qfct}) can be related to an earlier empirical QCD-inspired result proposed by Hagedorn \cite{hag83}
for high-energy pp and $\mathrm{p\bar{p}}$ collisions 
\begin{equation}
E\frac{\mathrm{d^3\sigma}}{\mathrm{d}p^3} = C\, (1 + p_\mathrm{T}/p_0)^{-n}
\label{hag}
\end{equation}
with a normalization constant $C$ and parameters $p_0, n$.
Choosing $p_0 = T/(q-1)$, $n = 1/(q-1)$ and replacing $p_\mathrm{T}$ by $m_\mathrm{T}$, Eqs.~(\ref{qfct}) and (\ref{hag}) are
found to be equivalent; see also Wilk and Wong \cite{wowi13} for pp. Hence, both expressions describe the transition from exponential ($\propto \exp(-m_\mathrm{T}/T)$ for $p_\mathrm{T}\rightarrow 0$ 
as in the J\"uttner distribution (\ref{juett})) to power-law behaviour
($\propto  (p_\mathrm{T}/nT)^{-n}$ for $p_\mathrm{T}\rightarrow \infty$). 

\begin{figure}[tph]
\includegraphics[width=8.4cm]{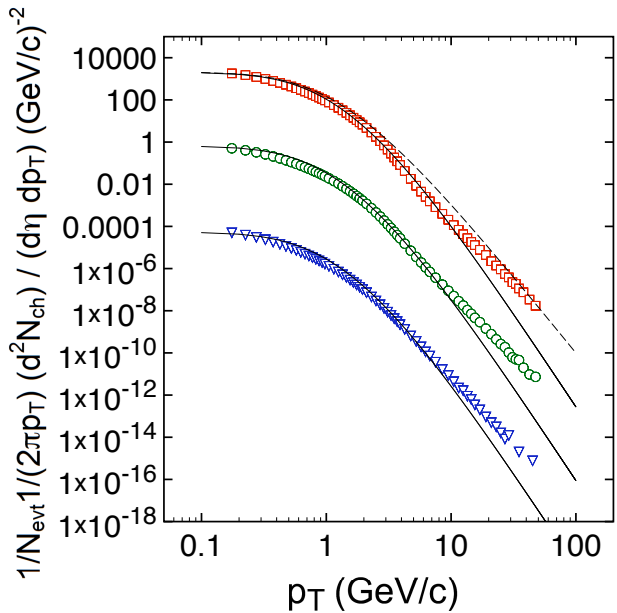}
\caption{(Color online) Transverse momentum distributions of produced charged hadrons in 2.76 TeV PbPb collisions calculated from $f(p_\mathrm{T})\propto [1 + (q-1)\, m_\mathrm{T}/T]^{1/(1-q)}$ compared with ALICE data \cite{abe13} for 0--5\%, 30--40\% and 70--80\% centralities (top to bottom).
Solid curves are for $q = 1.10$, the dashed curve is for $q = 1.12$. Peripheral spectra are scaled for better visibility, see Fig.~1 for absolute values.}
\label{fig2}
\end{figure}

Using Eq.~(\ref{qfct}) -- or equivalently, Eq.~(\ref{hag}) --, Fig.~\ref{fig2} shows calculated $p_\mathrm{T}$-distributions of produced charged hadrons for three centralities in 2.76 TeV PbPb compared with ALICE data from \cite{abe13} (peripheral spectra are scaled for better visibility, see Fig.~\ref{fig1} for absolute values; statistical and systematic error bars are smaller than the symbol size). Here the freezeout-temperature is $T \equiv T_\mathrm{F} = 120$ MeV and the average mass is
$m \equiv \langle m \rangle = 0.22$ GeV/$c^2$, as in Fig.~\ref{fig1}. 

The data are well represented through many orders of magnitude with $q = 1.10$ (Fig.~\ref{fig2}), but above $p_\mathrm{T} \sim 7$ GeV/$c$ deviations occur which are
attributed to hard processes that require a pQCD treatment. This limiting value of $p_\mathrm{T} \sim 7$ GeV/$c$ corresponds to a minimum in the nuclear modification factor for produced charged hadrons as function of $p_\mathrm{T}$ found in \cite{abe13}. 

Better results for the high-momentum tails would be achieved with slightly larger values of $q$ such 
as $q = 1.12$ corresponding to a power index $n = 8.3$ shown in Fig.~\ref{fig2}, but only at the expense of an unsatisfactory fit at mid-$p_\mathrm{T}$ values. 
It thus appears that apart from hard processes that can not be treated in a statistical approach, the functions (\ref{qfct}), (\ref{hag}) properly account for the transition from exponential to power-law spectra seen in the measured $p_T$-distributions .


Several authors have argued that distributions with $q > 1$ may reflect nonequilibrium systems with inhomogeneous
temperature and long-range interactions, e.g. \cite{alb00}. Previously Tsallis \cite{tsa88} 
had constructed a (``nonextensive'') $q \ne 1$  statistics that
incorporates Eq.~(\ref{qfct}) and would only 
in the absence of correlations assume the Boltzmann form -- see, however,
Balian and Nauenberg \cite{bana06} for a critical discussion of this view. 

There is presently no convincing theoretical derivation of the
value of $q$ -- or alternatively, of $n$ -- that is needed to reproduce the experimental $p_\mathrm{T}$-distributions in relativistic heavy-ion collisions. It is therefore not obvious 
from the present analysis what fraction of low-$p_\mathrm{T}$ particles is due to nonequilibrium processes 
that differ from thermal emission out of a single expanding fireball. In particular, one can not distinguish particles emitted from the
fireball and those arising from the fragmentation sources at low $p_\mathrm{T}$. Hence the analysis of transverse momentum distributions in terms of $q$-spectra 
is presently only suitable to distinguish high-$p_\mathrm{T}$ hard events from the bulk of (thermal and nonequilibrium) charged-hadron emission.

\section{Fragmentation distributions}

The distinction of particles emitted from the fireball and those from the fragmentation sources is more transparent
in rapidity or pseudorapidity distributions of produced charged hadrons. The existence of the fragmentation sources is 
evident from the measurements of stopping in heavy-ion collisions: Net-proton (proton minus antiproton) rapidity distributions $\mathrm{d}N_\mathrm{{p-\bar{p}}}/\mathrm{d}y$
exhibit two fragmentation peaks which are strongly overlapping at energies per particle pair of $\sqrt{s_{NN}} \lesssim 20$ GeV, but move apart 
at higher c.m. energies, leaving a midrapidity valley \cite{bjo83} at RHIC energies of 200 GeV \cite{bea04} that is predicted to broaden further at LHC energies
\cite{mtw09,dur14}. It is then largely depleted of baryons, with fragmentation peaks occuring in the rapidity regions $y = \mp~5-7$. Stopping is a highly nonequilibrium
process which is not suitable for any kind of thermal or equilibrium description.

The fragmentation peaks in stopping occur mainly due to the interaction of valence quarks with soft gluons in the respective other nucleus.
Their positions in rapidity space can be obtained from \cite{mtw09}
\begin{equation}
\frac{\mathrm{d}N_\mathrm{p-\bar{p}}}{\mathrm{d}y}=\frac{C}{(2\pi)^2}\int\frac{\mathrm{d}^2p_\mathrm{T}}{p_\mathrm{T}^2}x_1q_v(x_1,p_\mathrm{T})f_g(x_2,p_\mathrm{T})
\label{frag}
\end{equation}
for the peak in the forward region, and a corresponding symmetric contribution for the peak in the backward region that is obtained by replacing $y\rightarrow -y$.
Here $x_1 = p_\mathrm{T}/\sqrt{s}\exp(y)$ and $x_2 = p_\mathrm{T}/\sqrt{s}\exp(-y)$ are the respective longitudinal momentum fractions carried
by the valence quark $v$ in the projectile that undergoes stopping and the soft gluon $g$ in the target. The valence-quark distribution function is
$q_v(x_1,p_\mathrm{T})$ and the gluon distribution $f_g(x_2,p_\mathrm{T})$ is the Fourier transform of the forward
dipole scattering amplitude $N(x_2,r_\mathrm{T})$ for a quark
dipole of transverse size $r_\mathrm{T}$. The normalization constant $C$ is adjusted such that the integral of Eq.\,(\ref{frag}) yields the total number of participant 
protons in net-proton distributions
or baryons in net-baryon distributions.

\begin{figure}[tph]
\includegraphics[width=6cm]{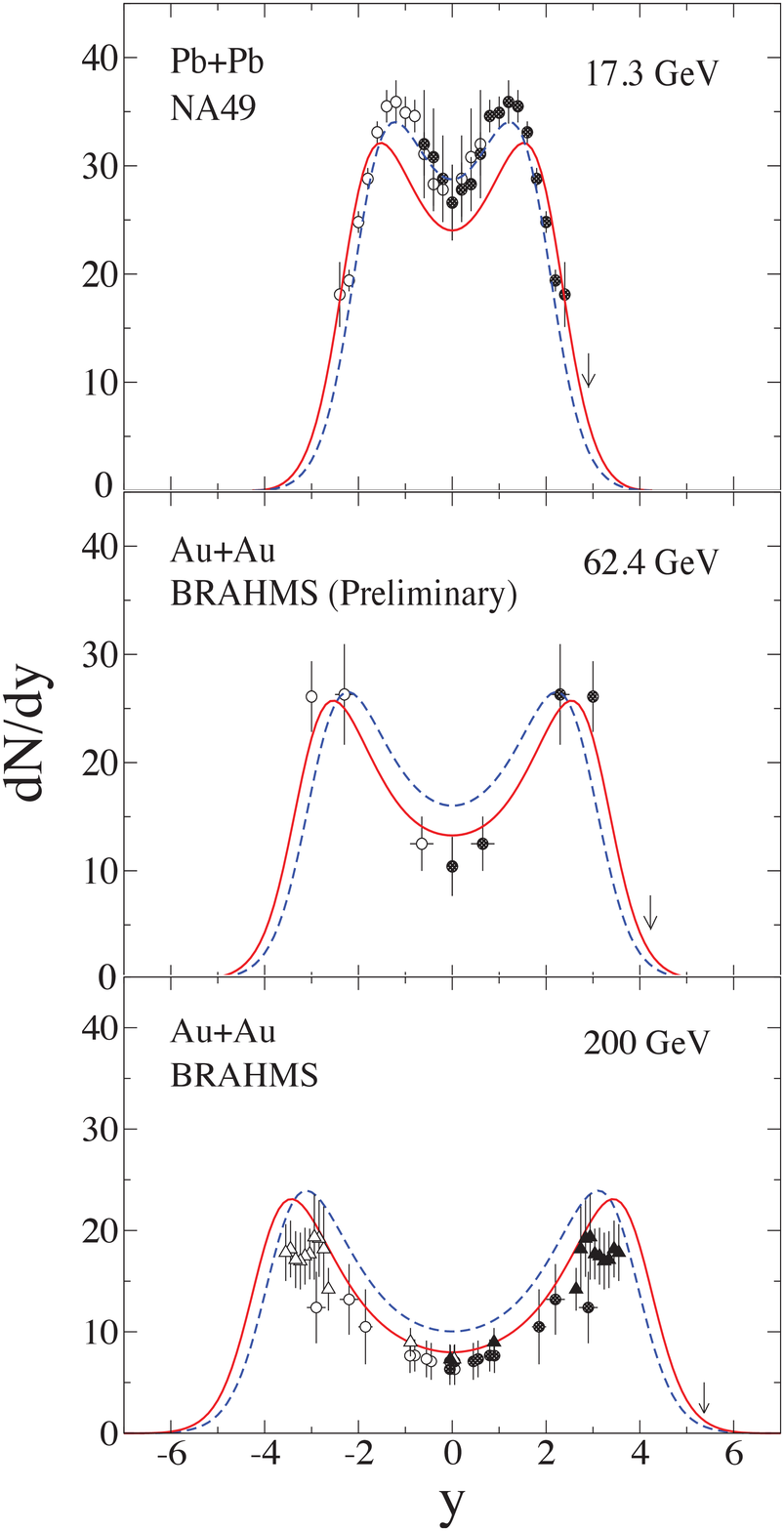}
\caption{\label{fig3}(Color online) Evidence for fragmentation sources: Rapidity distributions of net protons in central PbPb collisions at SPS energies of
$\sqrt{s_{NN}}$ = 17.3 GeV (top frame) compared with
NA49 data \cite{app99}. Solid curves correspond to a gluon saturation momentum $Q_\mathrm{s} = 0.9$ GeV/$c$ at $x = 0.01$, 
 dashed curves to $Q_\mathrm{s} = 1.2$ GeV/$c$.
 At RHIC
energies of 62.4 GeV (middle frame) and 200 GeV
(bottom frame) for central AuAu, theoretical
results are compared with BRAHMS net proton data~\cite{bea04}. The fragmentation peaks
move apart in rapidity space with increasing energy. Arrows indicate the
beam rapidities. From Mehtar-Tani and Wolschin \cite{mtw09,mtw15}.}
\end{figure}
 
The fragmentation peak positions $y_{\text{peak}}$ in rapidity space are at suffiently high energy -- in particular, at LHC energies --
 indicators for the gluon saturation scale
\begin{equation}
Q_\mathrm{s}^2 = A^{1/3}Q_0^2x^{-\lambda}
\end{equation}
with the mass number $A$, the momentum scale $Q_0$, the momentum fraction $x<1$  carried by the gluon and
the saturation-scale exponent $\lambda$. 

Rapidity distributions $\mathrm{d}N_\mathrm{{p-\bar{p}}}/\mathrm{d}y$ at SPS and RHIC energies are calculated within
the model of \cite{mtw09} for two values of the gluon saturation scale 
and compared with net-proton data from SPS and RHIC in Fig.~\ref{fig3} \cite{mtw15}.
A larger gluon saturation momentum $Q_\mathrm{s}$ produces more stopping, as does a larger mass
number $A$.
In the context of an investigation of particle production, the agreement between the calculated
stopping distributions and the data is taken as evidence for
the importance of fragmentation contributions also in charged-hadron production.

The peak positions are found to depend in a large c.m. energy range 
6.3 GeV $\leq \sqrt{s_{NN}} \leq 200$ GeV linearly
on the beam rapidity $y_{\text{beam}}$ and the saturation-scale exponent $\lambda$
according to \cite{mtw11}
\begin{equation}
y_{\text{peak}}=\frac{1}{1+\lambda}(y_{\text{beam}} - \ln A^{1/6})+ \text{const}
\end{equation}
and hence, at the current LHC energy of 5.02 TeV PbPb corresponding to $y_{\text{beam}} = \mp \ln({\sqrt{s_{NN}}/m_\mathrm{p}}) = \mp~8.586$  and with a gluon saturation-scale exponent $\lambda \sim 0.2$ one expects $y_{\text{peak}}\simeq \mp~6$.

Unfortunately the rapidity region of the peaks at LHC energies will therefore not be 
accessible for identified protons in the coming years due to the lack of
a suitable forward spectrometer at LHC. However, the partonic processes that mediate stopping also
contribute to hadron production and hence, one expects fragmentation events in particle production,
albeit with peaks occuring at somewhat smaller absolute rapidities than the ones for stopping.

Whereas in net-baryon (proton) distributions charged baryons produced from the gluonic source cancel out
because particles and antiparticles are generated in equal amounts, this is obviously not the case in charged-hadron
distributions. Here at sufficiently high energy  $\sqrt{s_{NN}} \gtrsim 20$ GeV three sources contribute and the dependence of their particle content on c.m. energy
differs: The fragmentation sources contain $N_\mathrm{ch}^{qg} \propto \ln(s_{NN}/s_0)$ charged hadrons;
the midrapidity-centered source that arises essentially from the interaction of low-$x$ gluons contains $N_\mathrm{ch}^{gg} \propto \ln^3(s_{NN}/s_0)$
charged hadrons, and becomes more important than the fragmentation sources at LHC energies \cite{gw15}.

Since the fragmentation distributions must exist in charged-hadron production because they can be
measured separately in net-proton data and the gluonic distribution
is known to be present in particle production, with particles and antiparticles 
produced in equal amounts,  
the total rapidity distribution for produced charged hadrons becomes
\begin{eqnarray}
\frac{\mathrm{d}N^{\mathrm{tot}}_{\mathrm{ch}}(y,t=\tau_{\mathrm{int}})}{\mathrm{d}y}=N_{\mathrm{ch}}^{{qg,1}}R_{1}(y,\tau_{\mathrm{int}}) + \\ \nonumber
N_{\mathrm{ch}}^{{gq,2}}R_{2}(y,\tau_{\mathrm{int}})+N_{\mathrm{ch}}^{gg}R_{gg}(y,\tau_{\mathrm{int}})
\label{normloc1}
\end{eqnarray}
with fragmentation distributions $R_{1,2}(y,t)$ and gluonic distributions $R_{gg}(y,t)$ calculated 
in a time-dependent phenomenological model such as the relativistic diffusion model (RDM) \cite{gw13}, or in microscopic theories. 
At the interaction time $t = \tau_\mathrm{int}$
the strong interaction ceases to act and theoretical distributions may be compared to data in a $\chi^2$-minimization.

In the relativistic diffusion model \cite{gw13}, the initial distribution functions are evolved up to $\tau_{int}/\tau_y$
with the rapidity relaxation time $\tau_y$ using the analytical moments equations. The mean values $\langle y_{1,2} \rangle$ 
of the fragmentation distributions that are related analytically to $\tau_{int}/\tau_y$ are determined 
from the data. The absolute
value of $\tau_{int}$ does not appear in this calculation because it would require a theory for $\tau_y$, which is 
not available to date. 

The three sources are evolved together, and the equilibration towards the thermal limit for both mean values and widths results from the 
evolution equation. The widths of the three sources at $\tau_{int}/\tau_y$  are, however, eventually determined
empirically in fits to the data because they implicitly include the effect of collective expansion and are therefore considerably larger
than the widths that may be calculated from the nonequilibrium evolution equation using the Einstein relations \cite{wols99}
and are also larger than the thermal limits for the widths. Hence the evolution equation is governing the statistical equilibration of the mean values of the three sources 
towards the thermal limit but the widths are empirically found to exceed the thermal values due to collective expansion.

In spite of its reasonable physical basis, the description of the nonequilibrium-statistical equilibration process based on three sources that evolve with time in rapidity space is a macroscopic idealization. This becomes especially evident when two of the three contributions become comparable, as it occurs e.g. in 5 TeV PbPb collisions at rapidity $y \simeq 4$\,: It seems not obvious why hadrons from valence quark-gluon (fragmentation) events should be out of equilibrium with respect to those from gluon-gluon events at any particular rapidity value. This is, however, different when considering the overall distribution of fragmentation and gluonic events in rapidity space and in particular, the time evolution of their mean values and widths: The nonequilibrium-statistical view should not be applied to individual events.

Since pseudorapidity distributions $\mathrm{d}N/\mathrm{d}\eta$ with $\eta = - \ln\,[\tan(\theta/2)]$ depend only on the scattering angle $\theta$ and
do not require particle identification, they are easier to obtain at large $\eta$-values (small scattering angles) compared to rapidity distributions at large 
values of $y = 0.5\,\ln[(E+p_{\parallel})/(E-p_{\parallel})]$.
To assess the significance of the fragmentation sources in particle production at LHC energies, it is therefore better to compare theoretical models
with pseudorapidity distributions of produced charged hadrons, rather than rapidity distributions of identified particles.

\section{Pseudorapidity distributions}

For produced charged hadrons in relativistic heavy-ion collisions, the pseudorapidity distributions $dN_\mathrm{ch}/d\eta$ thus
emerge from a superposition of the fragmentation sources and a midrapidity source that
is essentially due to low-$x$ gluons and rises rapidly in particle content according to $N_{{gg} }\propto \ln^3(s_{NN}/s_0)$ \cite{gw15}. 
The Jacobian that accounts for the conversion of rapidity distributions $\mathrm{\mathrm{d}N}_\mathrm{ch}/\mathrm{\mathrm{d}y}$ obtained in any theoretical
model to pseudorapidity distributions $\mathrm{\mathrm{d}N}_\mathrm{ch}/\mathrm{d}\eta$ 
can be calculated as
\begin{equation}
\frac{\mathrm{d}N}{\mathrm{d}\eta}=\frac{\mathrm{d}N}{\mathrm{d}y}\frac{\mathrm{d}y}{\mathrm{d}\eta}=
J(\eta,  m / p_\mathrm{T})\frac{\mathrm{d}N}{\mathrm{d}y}, 
\label{deta}
\end{equation}
\begin{equation}
{J(\eta,m /p_\mathrm{T})=\cosh({\eta})\cdot }
[1+(m/ p_\mathrm{T})^{2}
+\sinh^{2}(\eta)]^{-1/2}
\label{jac}
\end{equation}
with the hadron mass $m$ and the transverse momentum $p_\mathrm{T}$. Rather than calculating the Jacobian 
for charged-hadron distributions with
an average mass $\langle m \rangle$ and an average transverse momentum $\langle p_\mathrm{T} \rangle$, it is more precise to fix
 the mass $m$ at the pion mass $m_{\pi}$,  and calculate a corresponding effective mean transverse momentum  from
$\langle p_{\text{T,eff}}\rangle = m_{\pi} J_{y=0} /(1-J_{y=0}^2)^{1/2}$  \cite{rgw12}.
In this expression the Jacobian $J_{y=0}$ at midrapidity is taken from experiment 
for pions, kaons and protons.

\begin{figure}[tph]
\includegraphics[width=8.6cm]{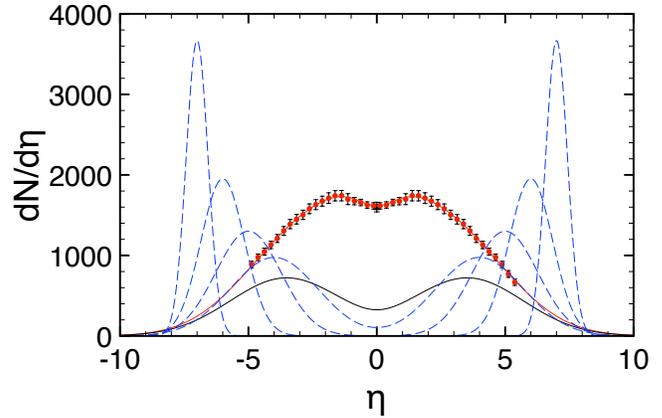}
\caption{\label{fig4}(Color online) Fragmentation sources for charged-hadron production in $\sqrt{s_{NN}}$ = 2.76 TeV PbPb collisions $(y_{\text{beam}} = \mp\,7.987)$.
Solid curves are from a $\chi^2$-minimization of analytical solutions in the relativistic diffusion model (RDM) \cite{gw13} with respect to
the ALICE data \cite{abel13} (upper curve including the gluonic source, lower curve fragmentation sources only). Dashed curves
indicate the time evolution of the fragmentation sources in the RDM. The fragmentation sources remain far from equilibrium at LHC energies.}
\end{figure}

The effective transverse momenta are smaller than the mean transverse momenta determined from the $p_\mathrm{T}$-distributions, and the corresponding effect of the Jacobian is therefore larger than that estimated
with $\langle p_\mathrm{T} \rangle$ taken from the transverse momentum distributions for each particle species. At high RHIC and LHC energies the effect of the Jacobian transformation remains, however, essentially confined to the midrapidity source.

There exist meanwhile several investigations that are considering or incorporating the relevance of the fragmentation sources in rapidity 
distributions of produced charged hadrons \cite{gw13,sami14,sah14,liu15,sark16}. In the 
relativistic diffusion model (RDM) \cite{wol99,wobi06}, the (pseudo-)rapidity distribution of produced particles emerges by construction of the model from an incoherent superposition of the  fragmentation components and a third source centered at (or near) midrapidity which is essentially due to low-$x$ gluon-gluon collisions. 
\begin{figure}[tph]
\begin{center}
\includegraphics[width=8.4cm]{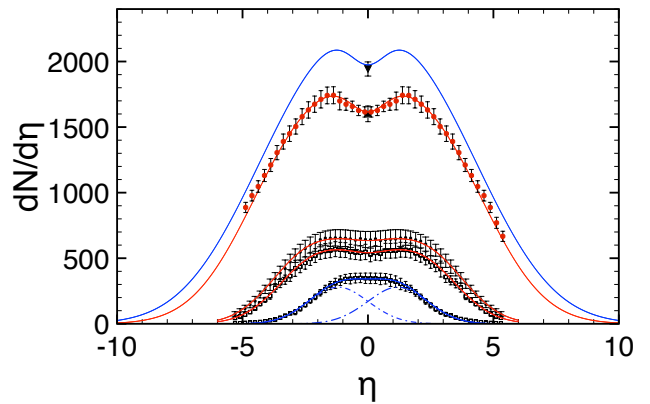}
\caption{\label{fig5}(Color online) The RDM pseudorapidity distribution functions for charged hadrons in central AuAu (RHIC) and PbPb (LHC) collisions
at c.m. energies of 19.6 GeV, 130 GeV, 200 GeV, 2.76 TeV shown here are optimized in $\chi^2$-fits with respect to the PHOBOS \cite{bb03,alv11} (bottom) and ALICE \cite{abel13} (top) data, with parameters from \cite{gw13}. The upper distribution function at 5.02 TeV is an extrapolation within the relativistic diffusion model. The 5.02 TeV midrapidity data point is from ALICE \cite{ad15}.}

\end{center}
\end{figure}

All three distribution functions (sources) $R_{1,2,{gg}}(y,t)$ evolve in time and are broadened in rapidity space as a consequence of diffusion-like processes governed by a Fokker-Planck equation (FPE). The fragmentation sources tend to shift towards midrapidity due to the drift term. Whereas this drift leads to a sizeable overlap of the fragmentation sources at lower (AGS, SPS) energies, their overlap at LHC energies is small due to the large rapidity gap and the very short interaction times, see Fig.~\ref{fig4} for the fragmentation sources in charged-hadron production from 0--5\% central 2.76 TeV PbPb collisions compared  with ALICE data \cite{abel13}. Here dashed curves
indicate the time evolution of the fragmentation sources.

The relevance of the fragmentation sources becomes particularly evident when investigating asymmetric systems such as 200 GeV dAu 
\cite{wobi06} or 5.02 TeV pPb  \cite{sgw15} where the total pseudorapidity distribution becomes asymmetric, is not centered at $\eta = 0$, and depends in a very distinctive manner on centrality. The distributions have steeper slopes in the p-going direction, and the midrapidity source shifts towards the Pb-going direction with increasing centrality. The agreement with the data enhances the credibility of the nonequilibrium three-sources approach. 

Hence, in the RDM the equilibration of the fragmentation sources with the gluonic source in the course of time is due to the nonequilibrium-statistical evolution equation  \cite{gw13}, with a thermal equilibrium distribution
emerging for $t\rightarrow \infty$. As is obvious from Fig.~\ref{fig4}, the charged-hadron distributions at LHC energies remain far from equilibrium. This is in spite of the observation that the three subdistributions are close to or have even reached local equilibrium, with an additional broadening due to collective expansion. 

In fact the phenomenological model of Liu et al. \cite{liu15} -- after its update from four to three sources according to the ones discussed here -- yields good fits of 
$\mathrm{d}N/\mathrm{d}\eta$-data in a large energy range from 19.6 GeV to 2.76 TeV with the assumption of local equilibrium in the three sources. 
There the midrapidity source is described in the Landau model \cite{lan53,lan55}.
Since the widths of the three sources have statistical and expansion contributions it is, however, difficult to determine from the data whether local equilibrium is actually reached in each source. 

This result relates to the current intense theoretical investigations of local equilibration within the gluonic source, e.g. \cite{larry14,raju14,fuku16} and references therein.
These works concern the microscopic equilibration mechanisms and eventually aim at at fully QCD-based nonperturbative description. A direct connection to the macroscopic 
investigation of equilibration among the three sources that is presented here is difficult to perform~conceptually and mathematically.
\begin{figure}[tph]
\includegraphics[width=8.8cm]{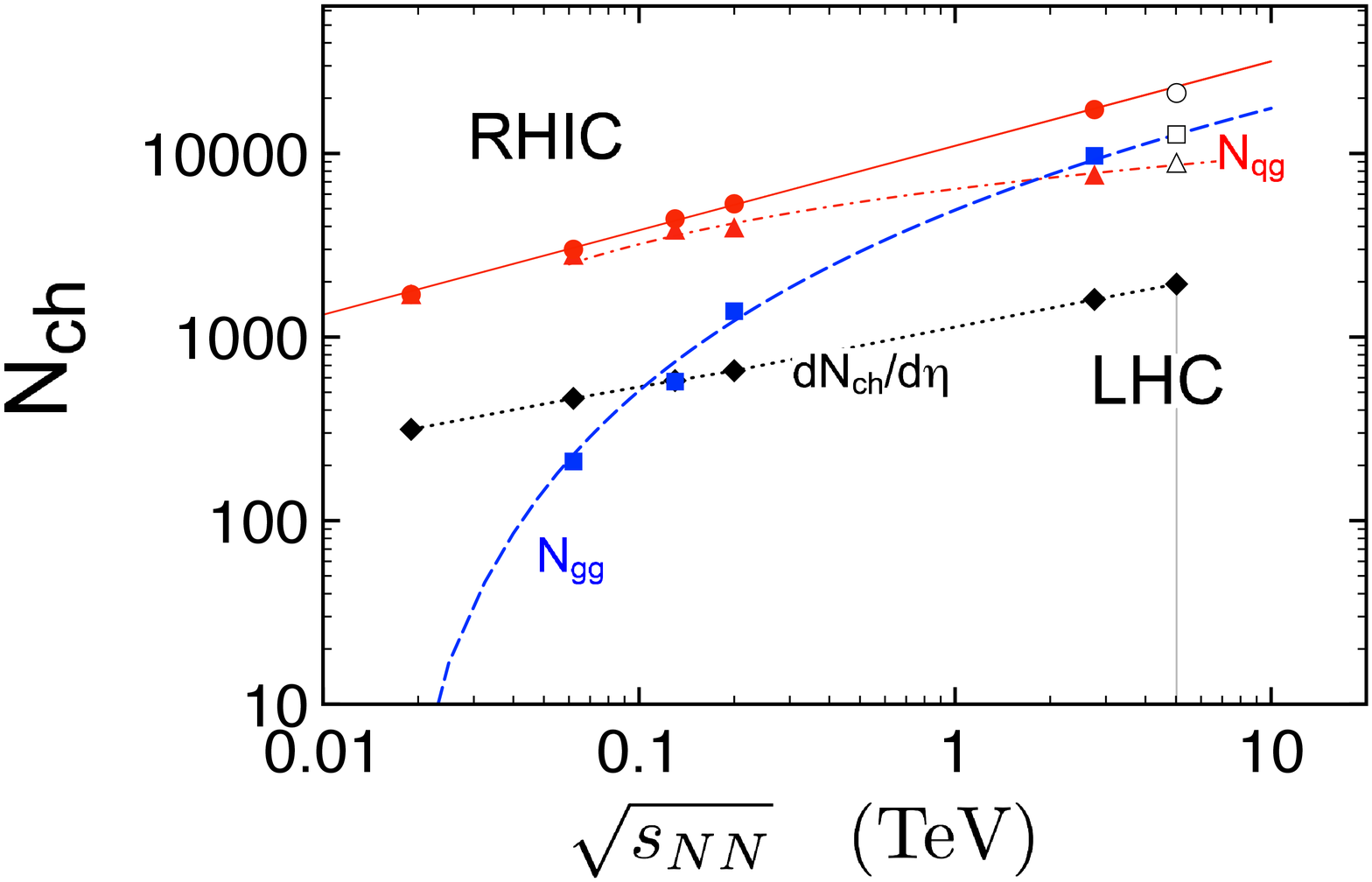}
\caption{\label{fig6}(Color online) The total charged-hadron production in central AuAu and PbPb collision in the energy region 19.6 GeV to 5.02 TeV is following a power law (solid upper line), 
whereas the particle content in the fragmentation sources is $N_\mathrm{qg} \propto \ln{(s_{NN}/s_0})$, dash-dotted curve. The particle content in the mid-rapidity source obeys $N_{gg} \propto \ln^3{(s_{NN}/s_0)}$,
dashed curve. The energy dependence of the measured mid-rapidity yields is shown as a dotted line, with PHOBOS data \cite{alv11} at RHIC energies,
and ALICE data \cite{aamo11,ad15} at 2.76  and 5.02 TeV. The vertical line indicates 5.02 TeV.}
\end{figure}

\begin{table}
\caption{\label{tab1}Three-sources RDM-parameters for charged-hadron production extrapolated to 5.02 TeV PbPb with $y_{\text{beam}} = \mp\,8.586$ at four centralities, see text. $\Gamma$ is the FWHM of the sources at the interaction time, $N_\mathrm{ch}$ the corresponding charged-particle content using the extrapolation formulae of \cite{gw15}. The last column gives the experimental midrapidity values from ALICE \cite{ad15}.}
\vspace{.3cm}
\begin{tabular}{lllllcr}
\hline\\
centrality&$\langle y_{1,2} \rangle$&$\Gamma_{1,2}$&$\Gamma_{gg}$&$N_\mathrm{ch}^{1+2}$&$N_\mathrm{ch}^{gg}$&$\frac{\mathrm{d}N}{\mathrm{d}\eta}|_{\eta \simeq 0}$\\
\hline\\
0--5\,\%&$\mp 3.5$&5.8&6.7&8644&12682&$1943\pm 54$\\
5--10\,\%&$\mp 3.5$&6.2&6.8&7623&10041&$1586\pm 46$\\
10--20\,\%&$\mp 3.5$&6.8&6.9&6023&7278&$1180\pm 31$\\
20--30\,\%&$\mp 3.5$&7.2&7.0&4271&4873&$786\pm 20$\\
\hline 
\end{tabular}
\end{table}

\begin{figure}[tph]
\includegraphics[width=8.8cm]{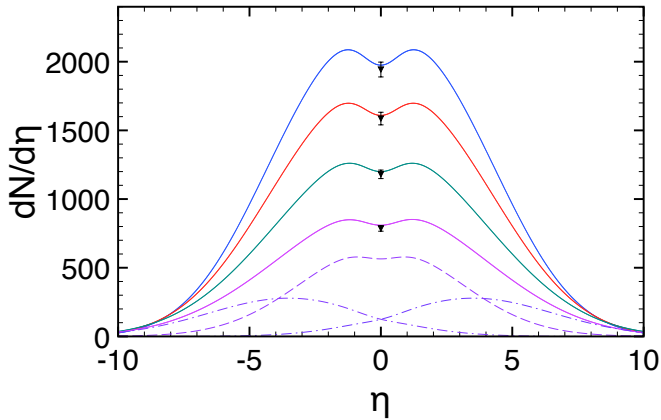}
\caption{(Color online) Pseudorapidity distributions for produced charged hadrons in 5.02 TeV PbPb collisions $(y_{\text{beam}} = \mp\,8.586)$ as functions of centrality, from bottom to top: 20--30\%, 10--20\%, 5--10\%, 0--5\%. Calculated RDM distributions (solid curves) are compared to midrapidity ALICE data from \cite{ad15}. For 20--30\% centrality fragmentation and gluonic distribution functions are shown separately.}
\label{fig7}
\end{figure}

The dependence of the pseudorapidity distributions on c.m. energy in central AuAu collisions at 19.6 GeV, 130 GeV, 200 GeV RHIC energies as well as 
in PbPb at 2.76 TeV and 5.02 TeV LHC energies is shown in Fig.~\ref{fig5}. In addition to RDM calculations with parameters for the lower energies from \cite{gw13}
compared with data from \cite{alv11,bb03,abe13}, an extrapolation to 5.02 TeV PbPb with parameters from Tab.~\ref{tab1}~ is compared with a recent
midrapidity ALICE data point at 0--5\% centrality \cite{ad15}. 

At the lowest RHIC energy of 19.6 GeV that is shown here -- which is comparable to the highest SPS energy in the upper frame of Fig.~\ref{fig3} -- only the fragmentation sources contribute (see dashed curves), but at higher energies the gluonic source rapidly catches up and
becomes the largest source of particle production at an energy of $\sim 2$ TeV, between energies reached at RHIC and LHC. 

The functional dependence of the particle content of the three sources on center-of-mass energy per particle pair $\sqrt{s_{NN}}$ has been investigated in \cite{gw15}. 
For $\sqrt{s_{NN}}\lesssim 20~$GeV the gluonic source is absent (19.6 GeV AuAu PHOBOS result in Fig.~\ref{fig5}) and charged-hadron production arises from the fragmentation sources which overlap in rapidity space and hence appear like a single gaussian (``thermal'') source. Experimentally the total charged-hadron production at these low energies has been found to depend linearly on $\ln(s_{NN}/s_0)$, see for example central PbPb NA50 data at 8.7 GeV and 17.3 GeV \cite{pri05} together with low-energy AuAu PHOBOS  results \cite{bb03}. 

In the RDM-analysis with three sources \cite{gw15} it turns out that the dependence of the fragmentation sources $N_\mathrm{ch}^{qg} \propto \ln(s_{NN}/s_0)$ indeed continues at higher energies up to 
the present maximum value for PbPb at 5.02 TeV, Fig.~\ref{fig6}. The gluonic source, however, has a much stronger energy dependence
$N_\mathrm{ch}^{gg} \propto \ln^3(s_{NN}/s_0)$ \cite{gw15}. The rise of the cross section in the central distribution is driven by the growth of the gluon density at small $x$ and theoretical arguments \cite{cheu11} suggest a ln$^2 s$ asymptotic behaviour that satisfies the Froissart bound \cite{fro61}. Since the beam rapidity is $\propto \ln(s_{NN})$, the integrated yield from the gluonic source then becomes proportional to ln$^3 s$. There exist also
further experimental confirmations of this result at RHIC energies based on STAR data for dijet production, see \cite{tom15} and references therein.

The sum of produced charged hadrons integrated over $\eta$ is then (accidentially) close to a power law $N_\mathrm{ch}^{\mathrm{tot}}\propto (s_{NN}/s_0)^{0.23}$ with $s_0 = 1$ TeV$^2$ as shown in Fig.~\ref{fig6} for central AuAu and PbPb collisions, upper line. At RHIC energies Busza noticed that the integrated charged-particle multiplicities scale as $\ln^2(s_{NN}/s_0)$
\cite{wbu04,wbu08}, but
the energy dependence up to LHC energies is found to be even stronger due to the high gluon density. The midrapidity yields for central AuAu and PbPb collisions
are 
\begin{equation}
\frac{dN_{ch}^{tot}}{d\eta}|_{\eta\simeq0}=1.15 \cdot 10^3 (s_{NN}/s_0)^{0.165}
\end{equation}
with $s_0=1~$TeV$^2$ (dotted line, data points from PHOBOS \cite{alv11} and ALICE \cite{aamo11,ad15}). 

 More detailed aspects of the interplay between fragmentation sources and gluonic source appear when investigating the centrality dependence of
 charged-hadron pseudorapidity distributions, as has been done in \cite{wobi06,sgw15} for the asymmetric systems 200 GeV dAu and 5.02 TeV pPb,
 and in \cite{gw13} for 2.76 TeV PbPb. For
 the newly investigated symmetric system 5.02 TeV PbPb charged-hadron distributions at centralities 20--30\%, 10--20\%, 5--10\% and  0--5\%
 are shown in Fig.~\ref{fig7}, with RDM-parameters in Tab.~\ref{tab1} extrapolated from the ones at lower energy in \cite{gw13}. 
 
 In a 0--5\% central collision, 
 about 20\% of the midrapidity yield still arises from the fragmentation sources, at 20--30\% centrality the fragmentation fraction at midrapidity is about 30\%.
 For 20--30\% centrality the three sources are shown separately in Fig.~\ref{fig7}.
 At all centralities, the system remains far from a thermalization of fragmentation sources and gluonic source: The three sources are separated
 in pseudorapidity space at freezeout.
 Although each of the sources is close to local equilibrium, thermalization would only be reached for very large times that are not accessible at these high energies.
 
 The RDM-extrapolations are seen to agree with the midrapidity data points recently measured by ALICE \cite{ad15}. Small modifications of the parameters may, however, be expected once $\eta$-dependent data become available.
 
\begin{figure}[tph]
\begin{center}
\includegraphics[width=4.6cm]{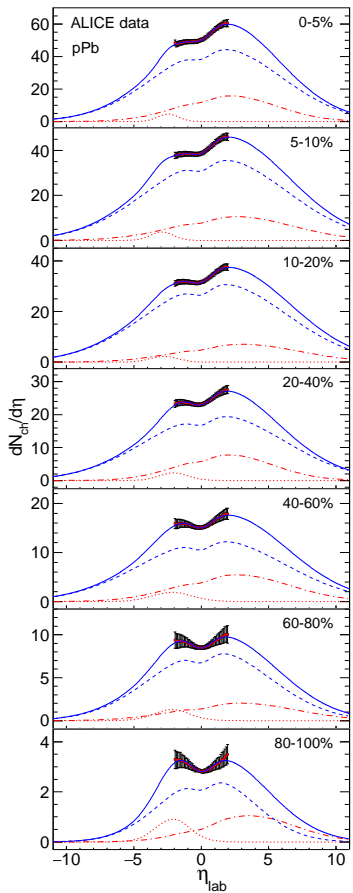}
\caption{\label{fig8}(Color online) The centrality-dependent RDM pseudorapidity distribution functions for charged hadrons in pPb collisions
at LHC c.m. energy of 5.02 TeV \cite{sgw15} are adjusted in the mid-rapidity region to the  ALICE data \cite{adam15} through $\chi^2$-minimizations.
The underlying distributions in the three--sources model are also shown, with the dashed curves
arising from gluon-gluon collisions, the dash-dotted curves from valence quark-gluon events in the Pb-going region ($y > 0$ in this plot), and the dotted curves in the p-going direction
(fragmentation sources) \cite{sgw16}.}
\end{center}
\end{figure}

 For the asymmetric system pPb at the same c.m. energy of 5.02 TeV,  pseudorapidity distributions of produced charged hadrons have been analyzed previously in the
 three-sources model at various centralities \cite{sgw15}. The calculated yields are higher in the Pb-going direction ($\eta > 0$ in this plot) than in the p-going direction, Fig.~\ref{fig8}. 
 
The underlying gluonic rapidity distributions are centered at the equilibrium values in the respective centrality bins which are calculated from energy-momentum conservation. The corresponding pseudorapidity distributions that are shown in the figure have a dip at midrapidity due
to the Jacobian transformation Eq.~(\ref{deta}) from rapidity to pseudorapidity space. The slopes of the tails depend on centrality, but they are always steeper on the proton-going  side. Particle creation from a gluon-dominated source, in addition to the
sources related to the valence part of the nucleons, had also been proposed by Bialas and
Czy\.z \cite{bia05}.
 
 A comparison ($\chi^2$-minimization) with the final ALICE data \cite{adam15} is shown in Fig.~\ref{fig8}, now with the additional constraint \cite{wobi06,sgw16} that the numbers of produced charged hadrons in the fragmentation sources are proportional to the numbers of participants. Again, the distribution functions remain far from thermal equilibrium at all centralities, they do not merge into a single thermal distribution.

The three-sources model is related to hydrodynamics and in particular, to viscous hydrodynamics which
also merely assumes local thermal equilibrium, with many 3D models \cite{gale13,koda16} being able to describe the 
pseudorapidity distributions of charged hadrons. To treat the stopping phase in the initial stage of the collision in
viscous hydrodynamics with the ensuing nonequilibrium distribution of baryon-rich matter one needs -- similarly to the RDM -- a three-sources model
with two counter-streaming fluids corresponding to the constituent nucleons of the projectile and target nuclei \cite{ams78,strott86}, and a third source (fireball)
in the midrapidity region that is associated with a fluid that is net-baryon free. Corresponding codes have been proposed \cite{ton06,iva16}
and implemented recently 
for AuAu collisions at low energies (BES\,II program at RHIC) with an emphasis on entropy production and
effective viscosity in a three-fluid
model using different equations of state for each of the sources with or without deconfinement transition. 
Such an approach may eventually also be applicable at the much higher LHC energies.

\section{Conclusion}
Deviations from thermal distribution functions for produced particles in relativistic heavy-ion collisions are sensitive indicators for nonequilibrium processes. These are discussed with special emphasis on the respective roles of fragmentation and central fireball sources in the longitudinal degrees of freedom that are quantified in pseudorapidity distributions of produced charged hadrons.

Transverse momentum distributions of produced charged hadrons in heavy-ion collisions at LHC energies are compatible with thermal Maxwell-Juettner distribution functions only at very small $p_\mathrm{T} \lesssim 0.5$ GeV/$c$. In a range $0.5$ GeV/$c \lesssim p_\mathrm{T} \lesssim 7$ GeV/$c$, collective expansion, thermal and nonequilibrium emission of particles contribute, but in a phenomenological model it is difficult to distinguish the respective contributions. 

Above $p_\mathrm{T} \sim 7$ GeV/$c$
hard processes are found to take over. These are not amenable to a statistical (equilibrium or nonequilibrium) description but require a pQCD treatment. Due to the rapid fall of
the transverse momentum distribution with $p_\mathrm{T}$, the contributions of hard and nonequilibrium processes to the total particle yield remain small when integrated
over $p_\mathrm{T}$, although they are relevant for the answer to the question whether the system is in thermal equilibrium. 

In the longitudinal degrees of freedom, net-proton rapidity distributions measured at SPS and RHIC energies clearly show the presence of the fragmentation sources in the stopping process.
Based on a QCD-inspired model that yields agreement with the data, these distributions result mainly from the interaction of the incoming valence quarks with low-$x$ gluons in the respective other nucleus. 

The fragmentation sources are present also in the production of charged hadrons, where they compete with the low-$x$ gluonic source that is centered at midrapidity
and rises rapidly in charged-particle content with the cube of the logarithmic c.m. energy.
Although all three sources are close to local equilibrium and broadened due to collective expansion, thermalization among them is not achieved during the short interaction time of $\sim$ 5--8~fm/$c$ in heavy-ion collisions at LHC energies. 

A three-sources nonequilibrium-statistical relativistic diffusion model that accounts for the time evolution of the fragmentation sources and the gluonic midrapidity source shows that the system remains far from thermal equilibrium among the sources, which would be reached only for large times $t\rightarrow\infty$. The model is used to predict pseudorapidity distributions and their centrality dependence for symmetric systems such as 5.02 TeV PbPb; it is also applicable for asymmetric systems such as pPb. 

In both cases and at all centralities the distribution functions remain far from thermal equilibrium distributions due to the large rapidity gap and the short interaction times of about $10^{-23}$ s at energies reached at the Relativistic Heavy-Ion Collider RHIC and the Large Hadron Collider LHC.
%
%
%
%
%
\begin{center}
\noindent
\bf{ACKNOWLEDGMENTS}
\end{center}
\rm
I am grateful to Jean-Paul Blaizot, Kenji Fukushima, Larry McLerran and Raju Venugopalan for conversations about local equilibration during their
stays at the Heidelberg Institute for Theoretical Physics, and to Frederike Vogel for her participation in the 5.02 TeV calculations. 
This work is supported by DFG through the Transregional Research Center TRR33 at the Universities of Bonn, LMU Munich and Heidelberg.
I thank my TR-colleagues for their cooperation.\\
\bibliography{gw_prc_nt}

\begin{thebibliography}{64}%
\makeatletter
\providecommand \@ifxundefined [1]{%
 \@ifx{#1\undefined}
}%
\providecommand \@ifnum [1]{%
 \ifnum #1\expandafter \@firstoftwo
 \else \expandafter \@secondoftwo
 \fi
}%
\providecommand \@ifx [1]{%
 \ifx #1\expandafter \@firstoftwo
 \else \expandafter \@secondoftwo
 \fi
}%
\providecommand \natexlab [1]{#1}%
\providecommand \enquote  [1]{``#1''}%
\providecommand \bibnamefont  [1]{#1}%
\providecommand \bibfnamefont [1]{#1}%
\providecommand \citenamefont [1]{#1}%
\providecommand \href@noop [0]{\@secondoftwo}%
\providecommand \href [0]{\begingroup \@sanitize@url \@href}%
\providecommand \@href[1]{\@@startlink{#1}\@@href}%
\providecommand \@@href[1]{\endgroup#1\@@endlink}%
\providecommand \@sanitize@url [0]{\catcode `\\12\catcode `\$12\catcode
  `\&12\catcode `\#12\catcode `\^12\catcode `\_12\catcode `\%12\relax}%
\providecommand \@@startlink[1]{}%
\providecommand \@@endlink[0]{}%
\providecommand \url  [0]{\begingroup\@sanitize@url \@url }%
\providecommand \@url [1]{\endgroup\@href {#1}{\urlprefix }}%
\providecommand \urlprefix  [0]{URL }%
\providecommand \Eprint [0]{\href }%
\providecommand \doibase [0]{http://dx.doi.org/}%
\providecommand \selectlanguage [0]{\@gobble}%
\providecommand \bibinfo  [0]{\@secondoftwo}%
\providecommand \bibfield  [0]{\@secondoftwo}%
\providecommand \translation [1]{[#1]}%
\providecommand \BibitemOpen [0]{}%
\providecommand \bibitemStop [0]{}%
\providecommand \bibitemNoStop [0]{.\EOS\space}%
\providecommand \EOS [0]{\spacefactor3000\relax}%
\providecommand \BibitemShut  [1]{\csname bibitem#1\endcsname}%
\let\auto@bib@innerbib\@empty
\bibitem [{\citenamefont {Hagedorn}(1965)}]{hag65}%
  \BibitemOpen
  \bibfield  {author} {\bibinfo {author} {\bibfnamefont {R.}~\bibnamefont
  {Hagedorn}},\ }\href@noop {} {\bibfield  {journal} {\bibinfo  {journal}
  {Nuovo Cim. Suppl.}\ }\textbf {\bibinfo {volume} {3}},\ \bibinfo {pages}
  {147} (\bibinfo {year} {1965})}\BibitemShut {NoStop}%
\bibitem [{\citenamefont {Braun-Munzinger}\ \emph {et~al.}(1995)\citenamefont
  {Braun-Munzinger}, \citenamefont {Stachel}, \citenamefont {Wessels},\ and\
  \citenamefont {Xu}}]{pbm95}%
  \BibitemOpen
  \bibfield  {author} {\bibinfo {author} {\bibfnamefont {P.}~\bibnamefont
  {Braun-Munzinger}}, \bibinfo {author} {\bibfnamefont {J.}~\bibnamefont
  {Stachel}}, \bibinfo {author} {\bibfnamefont {J.}~\bibnamefont {Wessels}}, \
  and\ \bibinfo {author} {\bibfnamefont {N.}~\bibnamefont {Xu}},\ }\href@noop
  {} {\bibfield  {journal} {\bibinfo  {journal} {Phys. Lett. B}\ }\textbf
  {\bibinfo {volume} {344}},\ \bibinfo {pages} {43} (\bibinfo {year}
  {1995})}\BibitemShut {NoStop}%
\bibitem [{\citenamefont {Manninen}\ and\ \citenamefont
  {Becattini}(2008)}]{mabe08}%
  \BibitemOpen
  \bibfield  {author} {\bibinfo {author} {\bibfnamefont {J.}~\bibnamefont
  {Manninen}}\ and\ \bibinfo {author} {\bibfnamefont {F.}~\bibnamefont
  {Becattini}},\ }\href@noop {} {\bibfield  {journal} {\bibinfo  {journal}
  {Phys. Rev. C}\ }\textbf {\bibinfo {volume} {78}},\ \bibinfo {pages} {054901}
  (\bibinfo {year} {2008})}\BibitemShut {NoStop}%
\bibitem [{\citenamefont {Braun-Munzinger}\ \emph {et~al.}(2016)\citenamefont
  {Braun-Munzinger}, \citenamefont {Koch}, \citenamefont {Schaefer},\ and\
  \citenamefont {Stachel}}]{pbm16}%
  \BibitemOpen
  \bibfield  {author} {\bibinfo {author} {\bibfnamefont {P.}~\bibnamefont
  {Braun-Munzinger}}, \bibinfo {author} {\bibfnamefont {V.}~\bibnamefont
  {Koch}}, \bibinfo {author} {\bibfnamefont {T.}~\bibnamefont {Schaefer}}, \
  and\ \bibinfo {author} {\bibfnamefont {J.}~\bibnamefont {Stachel}},\
  }\href@noop {} {\bibfield  {journal} {\bibinfo  {journal} {Phys. Rep.}\
  }\textbf {\bibinfo {volume} {621}},\ \bibinfo {pages} {76} (\bibinfo {year}
  {2016})}\BibitemShut {NoStop}%
\bibitem [{\citenamefont {Mather}\ \emph {et~al.}(1990)\citenamefont {Mather}
  \emph {et~al.}}]{cobe90}%
  \BibitemOpen
  \bibfield  {author} {\bibinfo {author} {\bibfnamefont {J.~C.}\ \bibnamefont
  {Mather}} \emph {et~al.} (\bibinfo {collaboration} {COBE Collaboration}),\
  }\href@noop {} {\bibfield  {journal} {\bibinfo  {journal} {Astrophys. J.}\
  }\textbf {\bibinfo {volume} {354}},\ \bibinfo {pages} {L37} (\bibinfo {year}
  {1990})}\BibitemShut {NoStop}%
\bibitem [{\citenamefont {Hinshaw}\ \emph {et~al.}(2009)\citenamefont {Hinshaw}
  \emph {et~al.}}]{wmap09}%
  \BibitemOpen
  \bibfield  {author} {\bibinfo {author} {\bibfnamefont {G.}~\bibnamefont
  {Hinshaw}} \emph {et~al.} (\bibinfo {collaboration} {WMAP Collaboration}),\
  }\href@noop {} {\bibfield  {journal} {\bibinfo  {journal} {Astrophys. J.
  Suppl.}\ }\textbf {\bibinfo {volume} {180}},\ \bibinfo {pages} {225}
  (\bibinfo {year} {2009})}\BibitemShut {NoStop}%
\bibitem [{\citenamefont {Adam}\ \emph
  {et~al.}(2016{\natexlab{a}})\citenamefont {Adam} \emph {et~al.}}]{planck15}%
  \BibitemOpen
  \bibfield  {author} {\bibinfo {author} {\bibfnamefont {R.}~\bibnamefont
  {Adam}} \emph {et~al.} (\bibinfo {collaboration} {Planck Collaboration}),\
  }\href@noop {} {\bibfield  {journal} {\bibinfo  {journal} {Astron.
  Astrophys., to be published}\ } (\bibinfo {year} {2016}{\natexlab{a}})},\
  \Eprint {http://arxiv.org/abs/1502.01582v2} {arXiv:1502.01582v2} \BibitemShut
  {NoStop}%
\bibitem [{\citenamefont {Abbas}\ \emph {et~al.}(2013)\citenamefont {Abbas}
  \emph {et~al.}}]{abb13}%
  \BibitemOpen
  \bibfield  {author} {\bibinfo {author} {\bibfnamefont {E.}~\bibnamefont
  {Abbas}} \emph {et~al.} (\bibinfo {collaboration} {ALICE Collaboration}),\
  }\href@noop {} {\bibfield  {journal} {\bibinfo  {journal} {Phys. Lett. B}\
  }\textbf {\bibinfo {volume} {726}},\ \bibinfo {pages} {610} (\bibinfo {year}
  {2013})}\BibitemShut {NoStop}%
\bibitem [{\citenamefont {Adam}\ \emph {et~al.}(2015)\citenamefont {Adam} \emph
  {et~al.}}]{adam15}%
  \BibitemOpen
  \bibfield  {author} {\bibinfo {author} {\bibfnamefont {J.}~\bibnamefont
  {Adam}} \emph {et~al.} (\bibinfo {collaboration} {ALICE Collaboration}),\
  }\href@noop {} {\bibfield  {journal} {\bibinfo  {journal} {Phys. Rev. C}\
  }\textbf {\bibinfo {volume} {91}},\ \bibinfo {pages} {064905} (\bibinfo
  {year} {2015})}\BibitemShut {NoStop}%
\bibitem [{\citenamefont {Cole}\ \emph {et~al.}(2013)\citenamefont {Cole} \emph
  {et~al.}}]{atlas13}%
  \BibitemOpen
  \bibfield  {author} {\bibinfo {author} {\bibfnamefont {B.}~\bibnamefont
  {Cole}} \emph {et~al.} (\bibinfo {collaboration} {ATLAS Collaboration}),\
  }\href@noop {} {\bibfield  {journal} {\bibinfo  {journal}
  {ATLAS-CONF-2013-096}\ } (\bibinfo {year} {2013})}\BibitemShut {NoStop}%
\bibitem [{\citenamefont {{B}earden~{\it et al.}
  (BRAHMS~Collaboration)}(2004)}]{bea04}%
  \BibitemOpen
  \bibfield  {author} {\bibinfo {author} {\bibfnamefont {I.~G.}\ \bibnamefont
  {{B}earden~{\it et al.} (BRAHMS~Collaboration)}},\ }\href@noop {} {\bibfield
  {journal} {\bibinfo  {journal} {Phys. Rev. Lett.}\ }\textbf {\bibinfo
  {volume} {93}},\ \bibinfo {pages} {102301} (\bibinfo {year}
  {2004})}\BibitemShut {NoStop}%
\bibitem [{\citenamefont {Mehtar-Tani}\ and\ \citenamefont
  {Wolschin}(2009)}]{mtw09}%
  \BibitemOpen
  \bibfield  {author} {\bibinfo {author} {\bibfnamefont {Y.}~\bibnamefont
  {Mehtar-Tani}}\ and\ \bibinfo {author} {\bibfnamefont {G.}~\bibnamefont
  {Wolschin}},\ }\href@noop {} {\bibfield  {journal} {\bibinfo  {journal}
  {Phys. Rev. Lett.}\ }\textbf {\bibinfo {volume} {102}},\ \bibinfo {pages}
  {182301} (\bibinfo {year} {2009})}\BibitemShut {NoStop}%
\bibitem [{\citenamefont {Dur{\~ a}es}\ \emph {et~al.}(2014)\citenamefont
  {Dur{\~ a}es}, \citenamefont {Giannini}, \citenamefont {Gon{\c c}alves},\
  and\ \citenamefont {Navarra}}]{dur14}%
  \BibitemOpen
  \bibfield  {author} {\bibinfo {author} {\bibfnamefont {F.~O.}\ \bibnamefont
  {Dur{\~ a}es}}, \bibinfo {author} {\bibfnamefont {A.~V.}\ \bibnamefont
  {Giannini}}, \bibinfo {author} {\bibfnamefont {V.~P.}\ \bibnamefont {Gon{\c
  c}alves}}, \ and\ \bibinfo {author} {\bibfnamefont {F.~S.}\ \bibnamefont
  {Navarra}},\ }\href@noop {} {\bibfield  {journal} {\bibinfo  {journal} {Phys.
  Rev. C}\ }\textbf {\bibinfo {volume} {89}},\ \bibinfo {pages} {035205}
  (\bibinfo {year} {2014})}\BibitemShut {NoStop}%
\bibitem [{\citenamefont {J{\"u}ttner}(1911)}]{juett11}%
  \BibitemOpen
  \bibfield  {author} {\bibinfo {author} {\bibfnamefont {F.}~\bibnamefont
  {J{\"u}ttner}},\ }\href@noop {} {\bibfield  {journal} {\bibinfo  {journal}
  {Annalen Phys.}\ }\textbf {\bibinfo {volume} {339}},\ \bibinfo {pages} {856}
  (\bibinfo {year} {1911})}\BibitemShut {NoStop}%
\bibitem [{\citenamefont {R{\"o}hrscheid}\ and\ \citenamefont
  {Wolschin}(2012)}]{rgw12}%
  \BibitemOpen
  \bibfield  {author} {\bibinfo {author} {\bibfnamefont {D.~M.}\ \bibnamefont
  {R{\"o}hrscheid}}\ and\ \bibinfo {author} {\bibfnamefont {G.}~\bibnamefont
  {Wolschin}},\ }\href@noop {} {\bibfield  {journal} {\bibinfo  {journal}
  {Phys. Rev. C}\ }\textbf {\bibinfo {volume} {86}},\ \bibinfo {pages} {024902}
  (\bibinfo {year} {2012})}\BibitemShut {NoStop}%
\bibitem [{\citenamefont {Kind}\ and\ \citenamefont {Wolschin}(2014)}]{kw14}%
  \BibitemOpen
  \bibfield  {author} {\bibinfo {author} {\bibfnamefont {T.}~\bibnamefont
  {Kind}}\ and\ \bibinfo {author} {\bibfnamefont {G.}~\bibnamefont
  {Wolschin}},\ }\href@noop {} {\bibfield  {journal} {\bibinfo  {journal}
  {priv. comm.}\ } (\bibinfo {year} {2014})}\BibitemShut {NoStop}%
\bibitem [{\citenamefont {Abelev}\ \emph
  {et~al.}(2013{\natexlab{a}})\citenamefont {Abelev} \emph {et~al.}}]{abe13}%
  \BibitemOpen
  \bibfield  {author} {\bibinfo {author} {\bibfnamefont {B.}~\bibnamefont
  {Abelev}} \emph {et~al.} (\bibinfo {collaboration} {ALICE Collaboration}),\
  }\href@noop {} {\bibfield  {journal} {\bibinfo  {journal} {Phys. Lett. B}\
  }\textbf {\bibinfo {volume} {720}},\ \bibinfo {pages} {52} (\bibinfo {year}
  {2013}{\natexlab{a}})}\BibitemShut {NoStop}%
\bibitem [{\citenamefont {{A}amodt}\ \emph {et~al.}(2011)\citenamefont
  {{A}amodt} \emph {et~al.}}]{aamo11}%
  \BibitemOpen
  \bibfield  {author} {\bibinfo {author} {\bibfnamefont {K.}~\bibnamefont
  {{A}amodt}} \emph {et~al.} (\bibinfo {collaboration} {ALICE Collaboration}),\
  }\href@noop {} {\bibfield  {journal} {\bibinfo  {journal} {Phys. Rev. Lett.}\
  }\textbf {\bibinfo {volume} {106}},\ \bibinfo {pages} {032301} (\bibinfo
  {year} {2011})}\BibitemShut {NoStop}%
\bibitem [{\citenamefont {Bearden}\ \emph {et~al.}(1997)\citenamefont {Bearden}
  \emph {et~al.}}]{bea78}%
  \BibitemOpen
  \bibfield  {author} {\bibinfo {author} {\bibfnamefont {I.~G.}\ \bibnamefont
  {Bearden}} \emph {et~al.} (\bibinfo {collaboration} {NA44 Collaboration}),\
  }\href@noop {} {\bibfield  {journal} {\bibinfo  {journal} {Phys. Rev. Lett.}\
  }\textbf {\bibinfo {volume} {78}},\ \bibinfo {pages} {2080} (\bibinfo {year}
  {1997})}\BibitemShut {NoStop}%
\bibitem [{\citenamefont {Siemens}\ and\ \citenamefont
  {Rasmussen}(1979)}]{sira79}%
  \BibitemOpen
  \bibfield  {author} {\bibinfo {author} {\bibfnamefont {P.~J.}\ \bibnamefont
  {Siemens}}\ and\ \bibinfo {author} {\bibfnamefont {J.~O.}\ \bibnamefont
  {Rasmussen}},\ }\href@noop {} {\bibfield  {journal} {\bibinfo  {journal}
  {Phys. Rev. Lett.}\ }\textbf {\bibinfo {volume} {42}},\ \bibinfo {pages}
  {880} (\bibinfo {year} {1979})}\BibitemShut {NoStop}%
\bibitem [{\citenamefont {Schnedermann}\ \emph {et~al.}(1993)\citenamefont
  {Schnedermann}, \citenamefont {Sollfrank},\ and\ \citenamefont
  {Heinz}}]{ssh93}%
  \BibitemOpen
  \bibfield  {author} {\bibinfo {author} {\bibfnamefont {E.}~\bibnamefont
  {Schnedermann}}, \bibinfo {author} {\bibfnamefont {J.}~\bibnamefont
  {Sollfrank}}, \ and\ \bibinfo {author} {\bibfnamefont {U.}~\bibnamefont
  {Heinz}},\ }\href@noop {} {\bibfield  {journal} {\bibinfo  {journal} {Phys.
  Rev. C}\ }\textbf {\bibinfo {volume} {48}},\ \bibinfo {pages} {2462}
  (\bibinfo {year} {1993})}\BibitemShut {NoStop}%
\bibitem [{\citenamefont {Heinz}\ and\ \citenamefont
  {Snellings}(2013)}]{hesne13}%
  \BibitemOpen
  \bibfield  {author} {\bibinfo {author} {\bibfnamefont {U.}~\bibnamefont
  {Heinz}}\ and\ \bibinfo {author} {\bibfnamefont {R.}~\bibnamefont
  {Snellings}},\ }\href@noop {} {\bibfield  {journal} {\bibinfo  {journal}
  {Annu. Rev. Nucl. Part. Sci.}\ }\textbf {\bibinfo {volume} {63}},\ \bibinfo
  {pages} {123} (\bibinfo {year} {2013})}\BibitemShut {NoStop}%
\bibitem [{\citenamefont {Gale}\ \emph {et~al.}(2013)\citenamefont {Gale},
  \citenamefont {Jeon},\ and\ \citenamefont {Schenke}}]{gale13}%
  \BibitemOpen
  \bibfield  {author} {\bibinfo {author} {\bibfnamefont {C.}~\bibnamefont
  {Gale}}, \bibinfo {author} {\bibfnamefont {S.}~\bibnamefont {Jeon}}, \ and\
  \bibinfo {author} {\bibfnamefont {B.}~\bibnamefont {Schenke}},\ }\href@noop
  {} {\bibfield  {journal} {\bibinfo  {journal} {Int. J. Mod. Phys. A}\
  }\textbf {\bibinfo {volume} {28}},\ \bibinfo {pages} {1340011} (\bibinfo
  {year} {2013})}\BibitemShut {NoStop}%
\bibitem [{\citenamefont {de~Souza}\ \emph {et~al.}(2016)\citenamefont
  {de~Souza}, \citenamefont {Koide},\ and\ \citenamefont {Kodama}}]{koda16}%
  \BibitemOpen
  \bibfield  {author} {\bibinfo {author} {\bibfnamefont {R.~D.}\ \bibnamefont
  {de~Souza}}, \bibinfo {author} {\bibfnamefont {T.}~\bibnamefont {Koide}}, \
  and\ \bibinfo {author} {\bibfnamefont {T.}~\bibnamefont {Kodama}},\
  }\href@noop {} {\bibfield  {journal} {\bibinfo  {journal} {Prog. Part. Nucl.
  Phys.}\ }\textbf {\bibinfo {volume} {86}},\ \bibinfo {pages} {35} (\bibinfo
  {year} {2016})}\BibitemShut {NoStop}%
\bibitem [{\citenamefont {Hagedorn}(1983)}]{hag83}%
  \BibitemOpen
  \bibfield  {author} {\bibinfo {author} {\bibfnamefont {R.}~\bibnamefont
  {Hagedorn}},\ }\href@noop {} {\bibfield  {journal} {\bibinfo  {journal} {Riv.
  Nuovo Cimento}\ }\textbf {\bibinfo {volume} {6}},\ \bibinfo {pages} {1}
  (\bibinfo {year} {1983})}\BibitemShut {NoStop}%
\bibitem [{\citenamefont {Wong}\ and\ \citenamefont {Wilk}(2013)}]{wowi13}%
  \BibitemOpen
  \bibfield  {author} {\bibinfo {author} {\bibfnamefont {C.~Y.}\ \bibnamefont
  {Wong}}\ and\ \bibinfo {author} {\bibfnamefont {G.}~\bibnamefont {Wilk}},\
  }\href@noop {} {\bibfield  {journal} {\bibinfo  {journal} {Phys. Rev. D}\
  }\textbf {\bibinfo {volume} {87}},\ \bibinfo {pages} {114007} (\bibinfo
  {year} {2013})}\BibitemShut {NoStop}%
\bibitem [{\citenamefont {Alberico}\ \emph {et~al.}(2000)\citenamefont
  {Alberico}, \citenamefont {Lavagno},\ and\ \citenamefont {Quarati}}]{alb00}%
  \BibitemOpen
  \bibfield  {author} {\bibinfo {author} {\bibfnamefont {W.~M.}\ \bibnamefont
  {Alberico}}, \bibinfo {author} {\bibfnamefont {A.}~\bibnamefont {Lavagno}}, \
  and\ \bibinfo {author} {\bibfnamefont {P.}~\bibnamefont {Quarati}},\
  }\href@noop {} {\bibfield  {journal} {\bibinfo  {journal} {Eur. Phys. J. C}\
  }\textbf {\bibinfo {volume} {12}},\ \bibinfo {pages} {499} (\bibinfo {year}
  {2000})}\BibitemShut {NoStop}%
\bibitem [{\citenamefont {Tsallis}(1988)}]{tsa88}%
  \BibitemOpen
  \bibfield  {author} {\bibinfo {author} {\bibfnamefont {C.}~\bibnamefont
  {Tsallis}},\ }\href@noop {} {\bibfield  {journal} {\bibinfo  {journal} {J.
  Stat. Phys.}\ }\textbf {\bibinfo {volume} {52}},\ \bibinfo {pages} {479}
  (\bibinfo {year} {1988})}\BibitemShut {NoStop}%
\bibitem [{\citenamefont {Balian}\ and\ \citenamefont
  {Nauenberg}(2006)}]{bana06}%
  \BibitemOpen
  \bibfield  {author} {\bibinfo {author} {\bibfnamefont {R.}~\bibnamefont
  {Balian}}\ and\ \bibinfo {author} {\bibfnamefont {M.}~\bibnamefont
  {Nauenberg}},\ }\href@noop {} {\bibfield  {journal} {\bibinfo  {journal}
  {Europhys. News}\ }\textbf {\bibinfo {volume} {37}},\ \bibinfo {pages} {9}
  (\bibinfo {year} {2006})}\BibitemShut {NoStop}%
\bibitem [{\citenamefont {Bjorken}(1983)}]{bjo83}%
  \BibitemOpen
  \bibfield  {author} {\bibinfo {author} {\bibfnamefont {J.~D.}\ \bibnamefont
  {Bjorken}},\ }\href@noop {} {\bibfield  {journal} {\bibinfo  {journal} {Phys.
  Rev. D}\ }\textbf {\bibinfo {volume} {27}},\ \bibinfo {pages} {140} (\bibinfo
  {year} {1983})}\BibitemShut {NoStop}%
\bibitem [{\citenamefont {{A}ppelsh{\"a}user~{\it et al.\/}
  (NA49~Collaboration)}(1999)}]{app99}%
  \BibitemOpen
  \bibfield  {author} {\bibinfo {author} {\bibfnamefont {H.}~\bibnamefont
  {{A}ppelsh{\"a}user~{\it et al.\/} (NA49~Collaboration)}},\ }\href@noop {}
  {\bibfield  {journal} {\bibinfo  {journal} {Phys. Rev. Lett.}\ }\textbf
  {\bibinfo {volume} {82}},\ \bibinfo {pages} {2471} (\bibinfo {year}
  {1999})}\BibitemShut {NoStop}%
\bibitem [{\citenamefont {Mehtar-Tani}\ and\ \citenamefont
  {Wolschin}(2015)}]{mtw15}%
  \BibitemOpen
  \bibfield  {author} {\bibinfo {author} {\bibfnamefont {Y.}~\bibnamefont
  {Mehtar-Tani}}\ and\ \bibinfo {author} {\bibfnamefont {G.}~\bibnamefont
  {Wolschin}},\ }\href@noop {} {\bibfield  {journal} {\bibinfo  {journal}
  {priv. comm.}\ } (\bibinfo {year} {2015})}\BibitemShut {NoStop}%
\bibitem [{\citenamefont {Mehtar-Tani}\ and\ \citenamefont
  {Wolschin}(2011)}]{mtw11}%
  \BibitemOpen
  \bibfield  {author} {\bibinfo {author} {\bibfnamefont {Y.}~\bibnamefont
  {Mehtar-Tani}}\ and\ \bibinfo {author} {\bibfnamefont {G.}~\bibnamefont
  {Wolschin}},\ }\href@noop {} {\bibfield  {journal} {\bibinfo  {journal}
  {Europhys. Lett.}\ }\textbf {\bibinfo {volume} {94}},\ \bibinfo {pages}
  {62003} (\bibinfo {year} {2011})}\BibitemShut {NoStop}%
\bibitem [{\citenamefont {Wolschin}(2015)}]{gw15}%
  \BibitemOpen
  \bibfield  {author} {\bibinfo {author} {\bibfnamefont {G.}~\bibnamefont
  {Wolschin}},\ }\href@noop {} {\bibfield  {journal} {\bibinfo  {journal}
  {Phys. Rev. C}\ }\textbf {\bibinfo {volume} {91}},\ \bibinfo {pages} {014905}
  (\bibinfo {year} {2015})}\BibitemShut {NoStop}%
\bibitem [{\citenamefont {Wolschin}(2013)}]{gw13}%
  \BibitemOpen
  \bibfield  {author} {\bibinfo {author} {\bibfnamefont {G.}~\bibnamefont
  {Wolschin}},\ }\href@noop {} {\bibfield  {journal} {\bibinfo  {journal} {J.
  Phys. G}\ }\textbf {\bibinfo {volume} {40}},\ \bibinfo {pages} {45104}
  (\bibinfo {year} {2013})}\BibitemShut {NoStop}%
\bibitem [{\citenamefont {Wolschin}(1999{\natexlab{a}})}]{wols99}%
  \BibitemOpen
  \bibfield  {author} {\bibinfo {author} {\bibfnamefont {G.}~\bibnamefont
  {Wolschin}},\ }\href@noop {} {\bibfield  {journal} {\bibinfo  {journal}
  {Europhys. Lett.}\ }\textbf {\bibinfo {volume} {47}},\ \bibinfo {pages} {30}
  (\bibinfo {year} {1999}{\natexlab{a}})}\BibitemShut {NoStop}%
\bibitem [{\citenamefont {Abelev}\ \emph
  {et~al.}(2013{\natexlab{b}})\citenamefont {Abelev} \emph {et~al.}}]{abel13}%
  \BibitemOpen
  \bibfield  {author} {\bibinfo {author} {\bibfnamefont {B.}~\bibnamefont
  {Abelev}} \emph {et~al.} (\bibinfo {collaboration} {ALICE Collaboration}),\
  }\href@noop {} {\bibfield  {journal} {\bibinfo  {journal} {Phys. Lett. B}\
  }\textbf {\bibinfo {volume} {726}},\ \bibinfo {pages} {610} (\bibinfo {year}
  {2013}{\natexlab{b}})}\BibitemShut {NoStop}%
\bibitem [{\citenamefont {Sahoo}\ and\ \citenamefont {Mishra}(2014)}]{sami14}%
  \BibitemOpen
  \bibfield  {author} {\bibinfo {author} {\bibfnamefont {R.}~\bibnamefont
  {Sahoo}}\ and\ \bibinfo {author} {\bibfnamefont {A.~N.}\ \bibnamefont
  {Mishra}},\ }\href@noop {} {\bibfield  {journal} {\bibinfo  {journal} {Int.
  J. Mod. Phys. E}\ }\textbf {\bibinfo {volume} {23}},\ \bibinfo {pages}
  {1450024} (\bibinfo {year} {2014})}\BibitemShut {NoStop}%
\bibitem [{\citenamefont {Sahoo}\ \emph {et~al.}(2015)\citenamefont {Sahoo},
  \citenamefont {Mishra}, \citenamefont {Behera},\ and\ \citenamefont
  {Nandi}}]{sah14}%
  \BibitemOpen
  \bibfield  {author} {\bibinfo {author} {\bibfnamefont {R.}~\bibnamefont
  {Sahoo}}, \bibinfo {author} {\bibfnamefont {A.~N.}\ \bibnamefont {Mishra}},
  \bibinfo {author} {\bibfnamefont {N.~K.}\ \bibnamefont {Behera}}, \ and\
  \bibinfo {author} {\bibfnamefont {B.~K.}\ \bibnamefont {Nandi}},\ }\href@noop
  {} {\bibfield  {journal} {\bibinfo  {journal} {Adv. High Energy Phys.}\
  }\textbf {\bibinfo {volume} {2015}},\ \bibinfo {pages} {612390} (\bibinfo
  {year} {2015})}\BibitemShut {NoStop}%
\bibitem [{\citenamefont {Gao}\ and\ \citenamefont {Liu}(2015)}]{liu15}%
  \BibitemOpen
  \bibfield  {author} {\bibinfo {author} {\bibfnamefont {L.-N.}\ \bibnamefont
  {Gao}}\ and\ \bibinfo {author} {\bibfnamefont {F.-H.}\ \bibnamefont {Liu}},\
  }\href@noop {} {\bibfield  {journal} {\bibinfo  {journal} {Adv. High Energy
  Phys.}\ }\textbf {\bibinfo {volume} {2015}},\ \bibinfo {pages} {184713}
  (\bibinfo {year} {2015})}\BibitemShut {NoStop}%
\bibitem [{\citenamefont {Sarkisyan}\ \emph {et~al.}(2016)\citenamefont
  {Sarkisyan}, \citenamefont {Mishra}, \citenamefont {Sahoo},\ and\
  \citenamefont {Sakharov}}]{sark16}%
  \BibitemOpen
  \bibfield  {author} {\bibinfo {author} {\bibfnamefont {E.~K.~G.}\
  \bibnamefont {Sarkisyan}}, \bibinfo {author} {\bibfnamefont {A.~N.}\
  \bibnamefont {Mishra}}, \bibinfo {author} {\bibfnamefont {R.}~\bibnamefont
  {Sahoo}}, \ and\ \bibinfo {author} {\bibfnamefont {A.~S.}\ \bibnamefont
  {Sakharov}},\ }\href@noop {} {\bibfield  {journal} {\bibinfo  {journal}
  {Phys. Rev. D}\ }\textbf {\bibinfo {volume} {93}},\ \bibinfo {pages} {054046}
  (\bibinfo {year} {2016})}\BibitemShut {NoStop}%
\bibitem [{\citenamefont {Wolschin}(1999{\natexlab{b}})}]{wol99}%
  \BibitemOpen
  \bibfield  {author} {\bibinfo {author} {\bibfnamefont {G.}~\bibnamefont
  {Wolschin}},\ }\href@noop {} {\bibfield  {journal} {\bibinfo  {journal} {Eur.
  Phys. J. A}\ }\textbf {\bibinfo {volume} {5}},\ \bibinfo {pages} {85}
  (\bibinfo {year} {1999}{\natexlab{b}})}\BibitemShut {NoStop}%
\bibitem [{\citenamefont {Wolschin}\ \emph {et~al.}(2006)\citenamefont
  {Wolschin}, \citenamefont {Biyajima}, \citenamefont {Mizoguchi},\ and\
  \citenamefont {Suzuki}}]{wobi06}%
  \BibitemOpen
  \bibfield  {author} {\bibinfo {author} {\bibfnamefont {G.}~\bibnamefont
  {Wolschin}}, \bibinfo {author} {\bibfnamefont {M.}~\bibnamefont {Biyajima}},
  \bibinfo {author} {\bibfnamefont {T.}~\bibnamefont {Mizoguchi}}, \ and\
  \bibinfo {author} {\bibfnamefont {N.}~\bibnamefont {Suzuki}},\ }\href@noop {}
  {\bibfield  {journal} {\bibinfo  {journal} {Phys. Lett. B}\ }\textbf
  {\bibinfo {volume} {633}},\ \bibinfo {pages} {38} (\bibinfo {year}
  {2006})}\BibitemShut {NoStop}%
\bibitem [{\citenamefont {{B}ack}\ \emph {et~al.}(2003)\citenamefont {{B}ack}
  \emph {et~al.}}]{bb03}%
  \BibitemOpen
  \bibfield  {author} {\bibinfo {author} {\bibfnamefont {B.~B.}\ \bibnamefont
  {{B}ack}} \emph {et~al.} (\bibinfo {collaboration} {PHOBOS Collaboration}),\
  }\href@noop {} {\bibfield  {journal} {\bibinfo  {journal} {Phys. Rev. Lett.}\
  }\textbf {\bibinfo {volume} {91}},\ \bibinfo {pages} {052303} (\bibinfo
  {year} {2003})}\BibitemShut {NoStop}%
\bibitem [{\citenamefont {{A}lver}\ \emph {et~al.}(2011)\citenamefont {{A}lver}
  \emph {et~al.}}]{alv11}%
  \BibitemOpen
  \bibfield  {author} {\bibinfo {author} {\bibfnamefont {B.}~\bibnamefont
  {{A}lver}} \emph {et~al.} (\bibinfo {collaboration} {PHOBOS Collaboration}),\
  }\href@noop {} {\bibfield  {journal} {\bibinfo  {journal} {Phys. Rev. C}\
  }\textbf {\bibinfo {volume} {83}},\ \bibinfo {pages} {024913} (\bibinfo
  {year} {2011})}\BibitemShut {NoStop}%
\bibitem [{\citenamefont {Adam}\ \emph
  {et~al.}(2016{\natexlab{b}})\citenamefont {Adam} \emph {et~al.}}]{ad15}%
  \BibitemOpen
  \bibfield  {author} {\bibinfo {author} {\bibfnamefont {J.}~\bibnamefont
  {Adam}} \emph {et~al.} (\bibinfo {collaboration} {ALICE Collaboration}),\
  }\href@noop {} {\bibfield  {journal} {\bibinfo  {journal} {Phys. Rev. Lett.}\
  }\textbf {\bibinfo {volume} {116}},\ \bibinfo {pages} {222302} (\bibinfo
  {year} {2016}{\natexlab{b}})}\BibitemShut {NoStop}%
\bibitem [{\citenamefont {Schulz}\ and\ \citenamefont
  {Wolschin}(2015)}]{sgw15}%
  \BibitemOpen
  \bibfield  {author} {\bibinfo {author} {\bibfnamefont {P.}~\bibnamefont
  {Schulz}}\ and\ \bibinfo {author} {\bibfnamefont {G.}~\bibnamefont
  {Wolschin}},\ }\href@noop {} {\bibfield  {journal} {\bibinfo  {journal} {Eur.
  Phys. J. A}\ }\textbf {\bibinfo {volume} {51}},\ \bibinfo {pages} {18}
  (\bibinfo {year} {2015})}\BibitemShut {NoStop}%
\bibitem [{\citenamefont {Landau}(1953)}]{lan53}%
  \BibitemOpen
  \bibfield  {author} {\bibinfo {author} {\bibfnamefont {L.~D.}\ \bibnamefont
  {Landau}},\ }\href@noop {} {\bibfield  {journal} {\bibinfo  {journal} {Izv.
  Akad. Nauk. Ser. Fiz.}\ }\textbf {\bibinfo {volume} {17}},\ \bibinfo {pages}
  {51} (\bibinfo {year} {1953})}\BibitemShut {NoStop}%
\bibitem [{\citenamefont {Belen'kji}\ and\ \citenamefont
  {Landau}(1955)}]{lan55}%
  \BibitemOpen
  \bibfield  {author} {\bibinfo {author} {\bibfnamefont {S.~Z.}\ \bibnamefont
  {Belen'kji}}\ and\ \bibinfo {author} {\bibfnamefont {L.~D.}\ \bibnamefont
  {Landau}},\ }\href@noop {} {\bibfield  {journal} {\bibinfo  {journal} {Usp.
  Fiz. Nauk.}\ }\textbf {\bibinfo {volume} {56}},\ \bibinfo {pages} {309}
  (\bibinfo {year} {1955})}\BibitemShut {NoStop}%
\bibitem [{\citenamefont {McLerran}(2014)}]{larry14}%
  \BibitemOpen
  \bibfield  {author} {\bibinfo {author} {\bibfnamefont {L.}~\bibnamefont
  {McLerran}},\ }\href@noop {} {\bibfield  {journal} {\bibinfo  {journal}
  {Nucl. Phys. A}\ }\textbf {\bibinfo {volume} {926}},\ \bibinfo {pages} {3}
  (\bibinfo {year} {2014})}\BibitemShut {NoStop}%
\bibitem [{\citenamefont {Venugopalan}(2014)}]{raju14}%
  \BibitemOpen
  \bibfield  {author} {\bibinfo {author} {\bibfnamefont {R.}~\bibnamefont
  {Venugopalan}},\ }\href@noop {} {\bibfield  {journal} {\bibinfo  {journal}
  {Nucl. Phys. A}\ }\textbf {\bibinfo {volume} {928}},\ \bibinfo {pages} {209}
  (\bibinfo {year} {2014})}\BibitemShut {NoStop}%
\bibitem [{\citenamefont {Fukushima}(2016)}]{fuku16}%
  \BibitemOpen
  \bibfield  {author} {\bibinfo {author} {\bibfnamefont {K.}~\bibnamefont
  {Fukushima}},\ }\href@noop {} {\bibfield  {journal} {\bibinfo  {journal}
  {Rep. Prog. Phys., to be published}\ } (\bibinfo {year} {2016})},\ \Eprint
  {http://arxiv.org/abs/1603.02340} {arXiv:1603.02340} \BibitemShut {NoStop}%
\bibitem [{\citenamefont {Prino}\ \emph {et~al.}(2005)\citenamefont {Prino}
  \emph {et~al.}}]{pri05}%
  \BibitemOpen
  \bibfield  {author} {\bibinfo {author} {\bibfnamefont {F.}~\bibnamefont
  {Prino}} \emph {et~al.} (\bibinfo {collaboration} {NA 50 Collaboration}),\
  }\href@noop {} {\bibfield  {journal} {\bibinfo  {journal} {J. Phys. Conf.
  Ser.}\ }\textbf {\bibinfo {volume} {5}},\ \bibinfo {pages} {86} (\bibinfo
  {year} {2005})}\BibitemShut {NoStop}%
\bibitem [{\citenamefont {Cheung}\ and\ \citenamefont {Chiu}(2011)}]{cheu11}%
  \BibitemOpen
  \bibfield  {author} {\bibinfo {author} {\bibfnamefont {M.-F.}\ \bibnamefont
  {Cheung}}\ and\ \bibinfo {author} {\bibfnamefont {C.~B.}\ \bibnamefont
  {Chiu}},\ }\href@noop {} {\bibfield  {journal} {\bibinfo  {journal}
  {arXiv:1111.6945}\ } (\bibinfo {year} {2011})}\BibitemShut {NoStop}%
\bibitem [{\citenamefont {Froissart}(1961)}]{fro61}%
  \BibitemOpen
  \bibfield  {author} {\bibinfo {author} {\bibfnamefont {M.}~\bibnamefont
  {Froissart}},\ }\href@noop {} {\bibfield  {journal} {\bibinfo  {journal}
  {Phys. Rev.}\ }\textbf {\bibinfo {volume} {123}},\ \bibinfo {pages} {1053}
  (\bibinfo {year} {1961})}\BibitemShut {NoStop}%
\bibitem [{\citenamefont {Trainor}\ and\ \citenamefont
  {Prindle}(2015)}]{tom15}%
  \BibitemOpen
  \bibfield  {author} {\bibinfo {author} {\bibfnamefont {T.}~\bibnamefont
  {Trainor}}\ and\ \bibinfo {author} {\bibfnamefont {D.~J.}\ \bibnamefont
  {Prindle}},\ }\href@noop {} {\  (\bibinfo {year} {2015})},\ \Eprint
  {http://arxiv.org/abs/1512.01599} {arXiv:1512.01599} \BibitemShut {NoStop}%
\bibitem [{\citenamefont {Busza}(2004)}]{wbu04}%
  \BibitemOpen
  \bibfield  {author} {\bibinfo {author} {\bibfnamefont {W.}~\bibnamefont
  {Busza}},\ }\href@noop {} {\bibfield  {journal} {\bibinfo  {journal} {Acta
  Phys. Polon. B}\ }\textbf {\bibinfo {volume} {35}},\ \bibinfo {pages} {2873}
  (\bibinfo {year} {2004})}\BibitemShut {NoStop}%
\bibitem [{\citenamefont {Busza}(2008)}]{wbu08}%
  \BibitemOpen
  \bibfield  {author} {\bibinfo {author} {\bibfnamefont {W.}~\bibnamefont
  {Busza}},\ }\href@noop {} {\bibfield  {journal} {\bibinfo  {journal} {J.
  Phys. G: Nucl. Part. Phys.}\ }\textbf {\bibinfo {volume} {35}},\ \bibinfo
  {pages} {044040} (\bibinfo {year} {2008})}\BibitemShut {NoStop}%
\bibitem [{\citenamefont {Bialas}\ and\ \citenamefont {Czy\.z}(2005)}]{bia05}%
  \BibitemOpen
  \bibfield  {author} {\bibinfo {author} {\bibfnamefont {A.}~\bibnamefont
  {Bialas}}\ and\ \bibinfo {author} {\bibfnamefont {W.}~\bibnamefont
  {Czy\.z}},\ }\href@noop {} {\bibfield  {journal} {\bibinfo  {journal} {Acta
  Phys. Polon. B}\ }\textbf {\bibinfo {volume} {36}},\ \bibinfo {pages} {905}
  (\bibinfo {year} {2005})}\BibitemShut {NoStop}%
\bibitem [{\citenamefont {Schulz}\ and\ \citenamefont
  {Wolschin}(2016)}]{sgw16}%
  \BibitemOpen
  \bibfield  {author} {\bibinfo {author} {\bibfnamefont {P.}~\bibnamefont
  {Schulz}}\ and\ \bibinfo {author} {\bibfnamefont {G.}~\bibnamefont
  {Wolschin}},\ }\href@noop {} {\bibfield  {journal} {\bibinfo  {journal}
  {priv. comm.}\ } (\bibinfo {year} {2016})}\BibitemShut {NoStop}%
\bibitem [{\citenamefont {Amsden}\ \emph {et~al.}(1978)\citenamefont {Amsden},
  \citenamefont {Goldhaber}, \citenamefont {Harlow},\ and\ \citenamefont
  {Nix}}]{ams78}%
  \BibitemOpen
  \bibfield  {author} {\bibinfo {author} {\bibfnamefont {A.~A.}\ \bibnamefont
  {Amsden}}, \bibinfo {author} {\bibfnamefont {A.~S.}\ \bibnamefont
  {Goldhaber}}, \bibinfo {author} {\bibfnamefont {F.~H.}\ \bibnamefont
  {Harlow}}, \ and\ \bibinfo {author} {\bibfnamefont {J.~R.}\ \bibnamefont
  {Nix}},\ }\href@noop {} {\bibfield  {journal} {\bibinfo  {journal} {Phys.
  Rev. C}\ }\textbf {\bibinfo {volume} {17}},\ \bibinfo {pages} {2080}
  (\bibinfo {year} {1978})}\BibitemShut {NoStop}%
\bibitem [{\citenamefont {Clare}\ and\ \citenamefont
  {Strottman}(1986)}]{strott86}%
  \BibitemOpen
  \bibfield  {author} {\bibinfo {author} {\bibfnamefont {R.~B.}\ \bibnamefont
  {Clare}}\ and\ \bibinfo {author} {\bibfnamefont {D.}~\bibnamefont
  {Strottman}},\ }\href@noop {} {\bibfield  {journal} {\bibinfo  {journal}
  {Phys. Rept.}\ }\textbf {\bibinfo {volume} {141}},\ \bibinfo {pages} {177}
  (\bibinfo {year} {1986})}\BibitemShut {NoStop}%
\bibitem [{\citenamefont {Ivanov}\ \emph {et~al.}(2006)\citenamefont {Ivanov},
  \citenamefont {Russkikh},\ and\ \citenamefont {Toneev}}]{ton06}%
  \BibitemOpen
  \bibfield  {author} {\bibinfo {author} {\bibfnamefont {Y.~B.}\ \bibnamefont
  {Ivanov}}, \bibinfo {author} {\bibfnamefont {V.~N.}\ \bibnamefont
  {Russkikh}}, \ and\ \bibinfo {author} {\bibfnamefont {V.~D.}\ \bibnamefont
  {Toneev}},\ }\href@noop {} {\bibfield  {journal} {\bibinfo  {journal} {Phys.
  Rev. C}\ }\textbf {\bibinfo {volume} {73}},\ \bibinfo {pages} {044904}
  (\bibinfo {year} {2006})}\BibitemShut {NoStop}%
\bibitem [{\citenamefont {Ivanov}\ and\ \citenamefont {Soldatov}()}]{iva16}%
  \BibitemOpen
  \bibfield  {author} {\bibinfo {author} {\bibfnamefont {Y.~B.}\ \bibnamefont
  {Ivanov}}\ and\ \bibinfo {author} {\bibfnamefont {A.~A.}\ \bibnamefont
  {Soldatov}},\ }\href@noop {} {\ }\Eprint {http://arxiv.org/abs/1605.0247}
  {arXiv:1605.0247} \BibitemShut {NoStop}%
\end{thebibliography}%
\end{document}